\documentclass[12pt,showpacs]{revtex4}

\usepackage{graphicx,graphics}
\usepackage{enumerate}
\usepackage{subfigure}
\usepackage{amsmath}
\usepackage{amsfonts}
\usepackage{amssymb}
\usepackage{amsthm}
\usepackage{psfrag}

\catcode`\@=11
\def\numberbysection{\@addtoreset{equation}{section}
\def\theequation{\arabic{section}.\arabic{equation}}}
\numberbysection

\def\be{\begin{equation}}
\def\ee{\end{equation}}
\def\bea{\begin{eqnarray}}
\def\eea{\end{eqnarray}}

\def\i{{\rm i}}
\newcommand\egal{&\!\!=\!\!&}
\newcommand\iden{&\!\!\equiv\!\!&}

\begin{document}

\title{Jamming probabilities for a vacancy in the dimer model}

\author{V.S. Poghosyan$^1$, V.B. Priezzhev$^1$ and P. Ruelle$^2$}
\affiliation{
$^1$Bogoliubov Laboratory of Theoretical Physics, JINR, 141980 Dubna, Russia\\
$^2$Institut de Physique Th\'{e}orique, Universit\'{e} catholique de Louvain, B-1348 Louvain-La-Neuve, Belgium}

\begin{abstract}
Following the recent proposal made by Bouttier {\it et al} [Phys. Rev. {\bf E} 76, 041140 (2007)], we study analytically the mobility properties of a single vacancy in the close-packed dimer model on the square lattice. Using the spanning web representation, we find determinantal expressions for various observable quantities. In the limiting case of large lattices, they can be reduced to the calculation of Toeplitz determinants and minors thereof. The probability for the vacancy to be strictly jammed and other diffusion characteristics are computed exactly.
\end{abstract}

\maketitle


\section{Introduction}

Everyone knows the popular child puzzle called ``15''.
It consists of a $4 \times 4$ grid of numbered sliding square tiles with one free slot (vacancy).
Initially the tiles are  jumbled.
In the course of the game, not only the empty slot moves around the grid,
but the (marked) tiles also move, and can take different relative positions for a fixed position of the vacancy.
The goal of the game is to unjumble the tiles by successive moves, each one consisting in sliding a tile into the empty slot.
It is easy to see, and this is true for a grid of arbitrary size, that the vacancy can freely reach an arbitrary position.

To complicate the game, take now rectangular tiles --dimers or dominoes-- of size $2\times 1$ or $1 \times 2$ instead of square ones,
and consider tiling a large grid with those, leaving, as before, an empty slot of unit surface, i.e. a monomer
(so the number of sites in the grid must be odd).
The possible movements of the vacancy are now conditioned by the orientation of the neighboring dimers,
since a dimer can slide only if it is oriented towards the vacancy.
The main question, raised and studied in \cite{Bouttier}, is then:
what are the positions that the vacancy can possibly reach under these moves ?

Bouttier {\it et al} \cite{Bouttier} have analyzed this question by looking at various quantities related to the statistics
of the domain accessible to the vacancy (unlike in the game ``15'', the dimers are not marked).
Building on the well-known Temperley correspondence between fully-packed dimer configurations
(i.e. with no vacancy) and spanning trees, they showed that the correspondence can be extended to dimer configurations with a single vacancy,
which are then put in bijection with spanning webs.
Generically, a spanning web consists of a central tree, growing from the site where the vacancy is located,
surrounded by a number of nested loops to which further branches are attached.
They proved that the domain accessible to the vacancy is exactly given by the tree component.
Given a uniform distribution on the dimer configurations for a fixed position of the vacancy,
the previous correspondence induces a distribution on the spanning webs with statistical weights depending on their number of loops, and implicitly defines a non-trivial distribution on the clusters of sites accessible to the vacancy.

Putting the vacancy at the central site of a square, odd-by-odd grid, Bouttier {\it et al} performed numerical calculations on finite grids and
then extrapolated their findings to very large or infinite lattices.
One of the main results of their analysis is that the delocalization probability ${\cal P}_L$, defined as the probability that the cluster of sites
accessible to the vacancy covers the whole grid, decays like $L^{-1/4}$ for large $L$, the linear size of the grid.
From this result, the probability $p(s)$ that the domain accessible to the vacancy has size $s$, in infinite volume,
was estimated to decay like $s^{-9/8}$ for large $s$.
At the other end of the scale, the values of $p_L(s)$ for small values of $s$ were computed exactly for finite grids,
and used to assess the asymptotic values in infinite volume.
The numerical values reported in \cite{Bouttier} for $p(1)$, the probability that the vacancy be strictly jammed, and $p(2)$,
the probability that the vacancy can make exactly one move, are
$p(1) = 0.107\,864\,376\,269\,049\,511\,98(1)$ and $p(2) = 0.055\,905\,353\,801\,942(1)$.
Moreover, using Plouffe's inverter \cite{plou} applied to the first ten digits of $1/\sqrt{p(1)}$,
the authors of \cite{Bouttier} were able to conjecture the exact value of $p(1)$, namely $p(1)=57/4-10\sqrt{2}$.
The exact value of $p(2)$ remained unknown.

Our aim of this paper is to revisit these questions, and to provide an analytical derivation of some of the results mentioned above,
mainly the values of $p(1)$ and $p(2)$, as well as other quantities, which are relevant to the diffusion properties of the vacancy.
In particular we obtain the exact value of $p(2)$,
\begin{equation}
p(2) = {1 \over 32}(72\,817 \sqrt{2} - 102\,977),
\end{equation}
which agrees with all significant digits given in \cite{Bouttier}.

The plan of the article is as follows.
In Section \ref{sec2} we review the bijection between dimer configurations with the fixed vacancy and spanning webs,
as formulated and developed in \cite{Bouttier}, and discuss the spanning web enumeration technique.
In Section \ref{sec3}, we discuss the situation when the spanning webs can include a maximal, finite number of loops,
in the infinite volume limit, and relate it to the finite-size scaling of the delocalization probability found in \cite{Bouttier}.
In Section \ref{sec4} we analyze the probability that the vacancy take at least one step in a particular direction,
while Section \ref{sec5} contains the results announced above for $p(1)$ and $p(2)$.


\section{The model and the spanning web representation}\label{sec2}

\begin{figure}[h!]
\includegraphics[width=160mm]{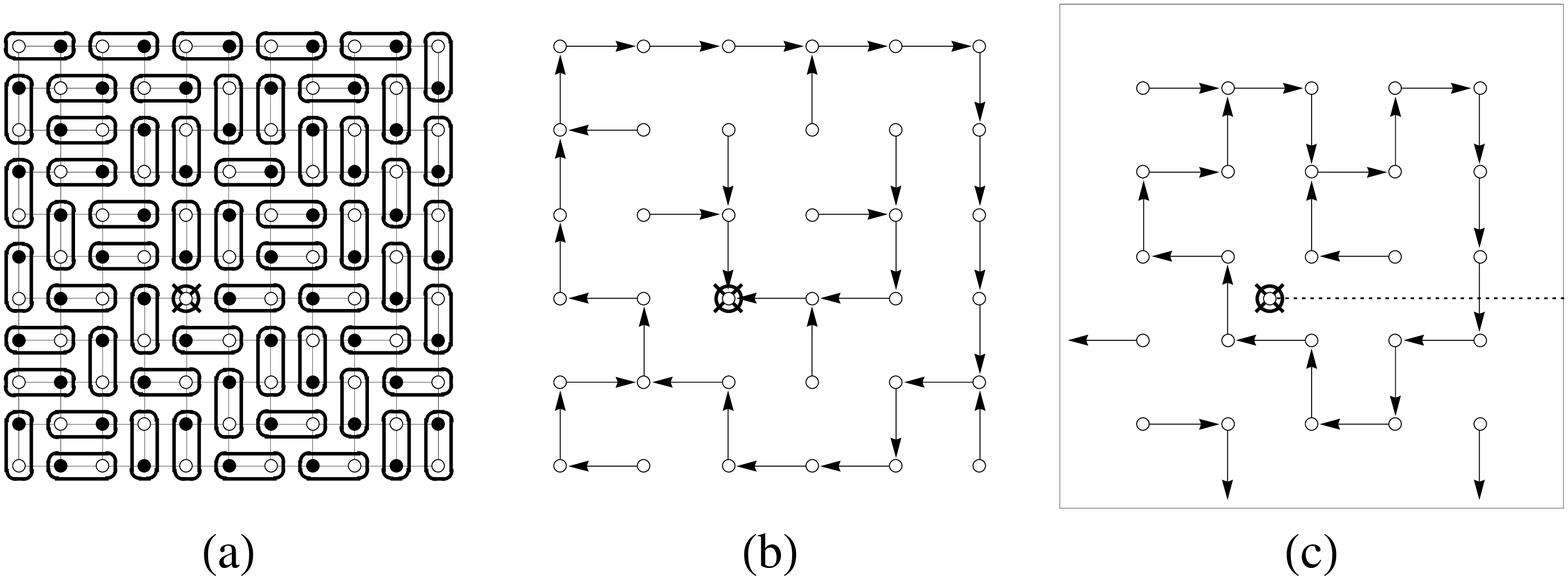}
\caption{(a) Example of dimer configuration with one odd vacancy (marked by $\otimes$) on a $11 \times 11$ grid.
The other two panels show the corresponding spanning webs defined (b) on the odd sublattice $\mathcal{L}_{\rm odd}$, and,
(c) on the even sublattice $\mathcal{L}_{\rm even}$.}
\label{fig1}
\end{figure}

Consider a square grid $\mathcal{L}$ of odd size.
Let the vertices be colored in black and white like a chessboard, so that neighboring vertices have different colors.
For definiteness, we assume that the corner vertices of the grid, which all have the same color, are white.
With this coloring, there is one more white vertex than black ones so that the grid can be fully covered by dimers
with a single vacancy on a white vertex.
Fig.\ref{fig1}a shows a possible configuration of dimers with the vacancy (represented as $\otimes$).

According to \cite{Bouttier}, a dimer lying next to the vacancy can move (by one site) and cover the vacancy if it is oriented towards it.
After the sliding of the dimer, the vacancy has moved by two sites, but remains on the white sublattice.
In the example of Fig.\ref{fig1}a, the upper and the right neighboring dimers of the vacancy can be moved; equivalently,
the vacancy can move two sites upwards or two sites rightwards.

The coordinates of each white vertex are either both even or both odd, so there are two types of white vertices, even and odd.
The sublattices of even and odd white vertices are called respectively even and odd as well,
and denoted $\mathcal{L}_{\rm even}$ and $\mathcal{L}_{\rm odd}$ ($\mathcal{L}_{\rm even}$ is slightly smaller than $\mathcal{L}_{\rm odd}$).
If the vacancy or a dimer are located on an even (odd) vertex, then we will call them respectively an even (odd) vacancy and an even (odd) dimer.

It is easy to see that the parity of the vacancy does not change during its motions.
Assume that it is odd, like in the example of Fig.\ref{fig1}a.
To set up the correspondence between dimer and spanning web configurations, we first remove the black and even vertices and keep the odd dimers only.
We then replace each odd dimer by an arrow of length $2$, starting from the odd vertex and directed along the dimer.
The arrow configuration obtained from the example in Fig.\ref{fig1}a is shown in Fig.\ref{fig1}b.
The set of all possible arrow configurations obtained in this way from dimer configurations with a fixed odd vacancy forms
the set of spanning webs defined on the odd sublattice.
As illustrated in Fig.\ref{fig1}b, a spanning web consists of a central tree component,
rooted at the vacancy location and oriented towards it, surrounded by a number of nested directed loops with branches attached to the loops.
The sites contained in the central tree are precisely the sites which are accessible to the vacancy.
Indeed, looking back at the original dimer configuration, one sees that the central tree is enclosed by a loop of even dimers
which acts as a cage for the vacancy.
If the central tree covers the whole of the odd sublattice, no loop is allowed and the spanning web is a spanning tree.
This is automatically the case when the vacancy is at a boundary site.

Conversely, if we have a fixed spanning web on the odd sublattice, we can reconstruct the dimer configuration up to the orientation
of the dimer loops on the even sublattice.
Indeed the set of even dimers also form a spanning web on the even sublattice,
which is dual to the one on the odd sublattice (see Fig.\ref{fig1}c).
The central piece of an even spanning web is a loop, whose interior is the domain accessible to the odd vacancy.
The even spanning web has the same number of loops as its dual odd web, but its tree components are oriented towards the exterior of the lattice.
An odd spanning web fixes the shape of the even spanning web but not the orientation of its loops (if any).
Since there are two possible orientations per loop, $2^n$ dimer configurations correspond to each odd (or even) spanning web with $n$ loops.
So we can enumerate all dimer configurations on the original lattice by counting the spanning webs on the odd (or even) sublattice with a weight $2^n$.

\begin{figure}[t]
\includegraphics[width=160mm]{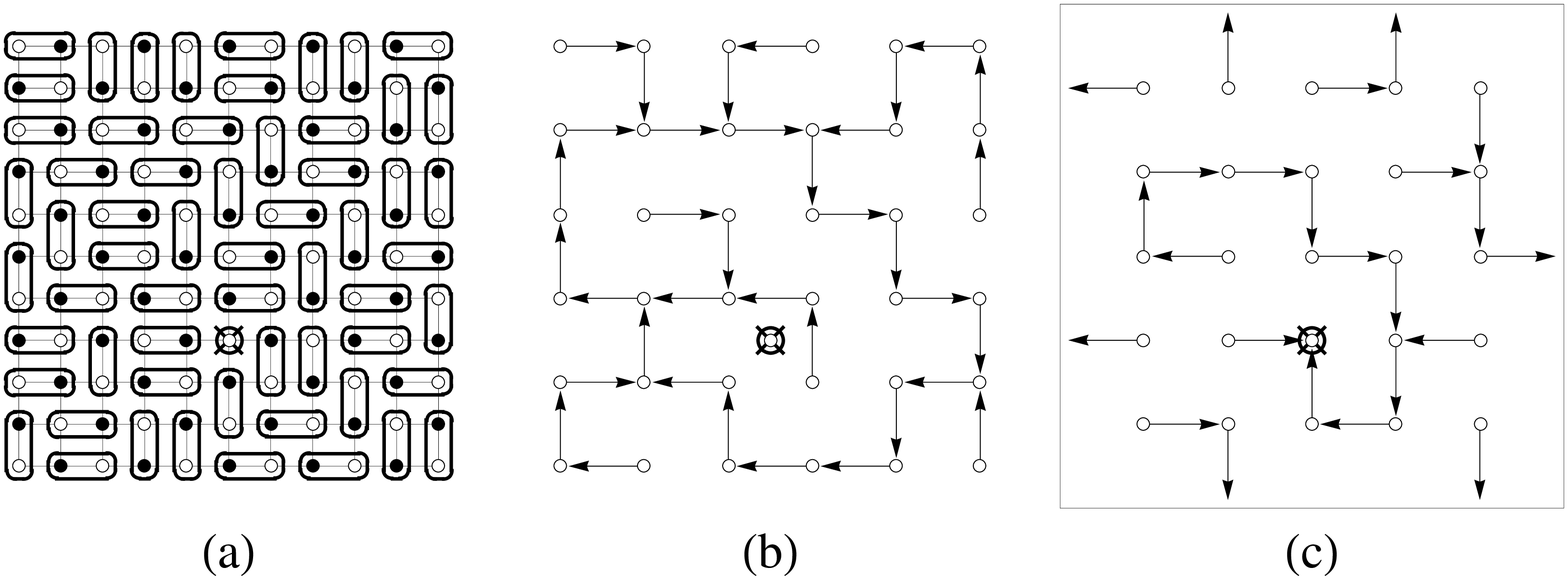}
\caption{
(a) Example of dimer configuration with even vacancy.
(b) Odd spanning web obtained from (a).
(c) Even spanning web obtained from (a).
}\label{fig2}
\end{figure}

The same construction can be made when the vacancy is even (see Fig.\ref{fig2}),
but in this case the situation is slightly asymmetric with respect to the choice of even or odd spanning webs.
Indeed it is not difficult to see that the odd spanning webs contain one more loop than the even spanning webs.
This affects the formulae with a factor $1/2$ or 2 depending on the choice one makes,
namely $Z_{\rm dimer} = {1 \over 2} Z_{\rm odd\,web} = 2 Z_{\rm even\,web}$.

Which spanning webs and vacancy, even or odd, we use is a matter of convenience.
In \cite{Bouttier}, the odd vacancy is located at the center of a $(4L+1) \times (4L+1)$ grid,
and the authors chose to work with the odd spanning webs.

We finish this section by recalling how the enumeration can be carried out.
We assume that the vacancy is odd, and we choose the spanning webs on the even sublattice $\mathcal{L}_{\rm even}$,
taken to be finite for the moment. As remarked above, the outmost piece of such a web is a (disconnected) tree,
rooted at the exterior of $\mathcal{L}_{\rm even}$. From Kirchhoff's theorem,
the number of spanning webs for which the outer tree fully covers $\mathcal{L}_{\rm even}$, i.e. spanning trees, is equal to
\begin{equation}
Z_{\rm tree} = \det \Delta^{\rm op},
\end{equation}
where $\Delta^{\rm op}$ is the discrete Laplacian on $\mathcal{L}_{\rm even}$ with open boundary conditions,
\begin{equation}
(\Delta^{\rm op})_{ij} = \begin{cases}
4 & \text{if $i=j$},\\
\noalign{\vspace{-3mm}}
-1 & \text{if $i,j$ are nearest neighbours}, \\
\noalign{\vspace{-3mm}}
0 & \text{otherwise}.
\end{cases}
\end{equation}

In order to count not only the spanning trees but also the spanning webs, we have to allow for loops circling around the vacancy
(which lies on $\mathcal{L}_{\rm odd}$).
This can be done by choosing a defect line connecting the vacancy to the boundary, for instance the dotted line shown in Fig.\ref{fig1}c.
Then we define a frustrated Laplacian matrix $\tilde\Delta^{\rm op}$, equal to $\Delta^{\rm op}$  except that
$(\tilde\Delta^{\rm op})_{ij}=+1$ for all those nearest neighbour sites connected by a bond (of $\mathcal{L}_{\rm even}$)
which crosses the defect line. Because any loop enclosing the vacancy necessarily contains an odd number of frustrated edges,
it can then be shown \cite{Bouttier,ipr} that the determinant of $\tilde\Delta^{\rm op}$ counts the number of spanning webs with the correct weight,
namely $2^n$ if the web contains $n$ loops,
\begin{equation}
Z_{\rm web} = \det \tilde\Delta^{\rm op}.
\end{equation}

If $(i_k,j_k)$ are the pairs of neighbouring sites separated by the defect line, for $k=1,2,\ldots,N$, where $N$ depends on the size of the grid, the position of the vacancy and the defect line that has been chosen, the matrix $\tilde\Delta^{\rm op}$ can be seen as a perturbation of $\Delta^{\rm op}$, $\tilde\Delta^{\rm op} = \Delta^{\rm op} + B$, with
\begin{equation}
B_{i_k,j_k} = B_{j_k,i_k} = 2, \qquad B_{ij} = 0 \quad {\rm elsewhere}.
\end{equation}
This allows to relate the number of spanning webs to the number of spanning trees,
\begin{equation}
{Z_{\rm web} \over Z_{\rm tree}} = \det(I + G^{\rm op} B),
\end{equation}
where $G^{\rm op}$ is the Green matrix, $G^{\rm op}=(\Delta^{\rm op})^{-1}$.
This number is the inverse of the localization probability introduced in \cite{Bouttier}.
Let us note that the previous formula requires that $Z_{\rm tree} = \det \Delta$ be non-zero, and so is not defined in the case of an even vacancy if the odd spanning webs are used, since $\Delta$ would be subjected to the closed boundary conditions, implying $\det \Delta^{\rm cl} = 0$.


\section{Finite defect lines}\label{sec3}

\begin{figure}[h!]
\includegraphics[width=100mm]{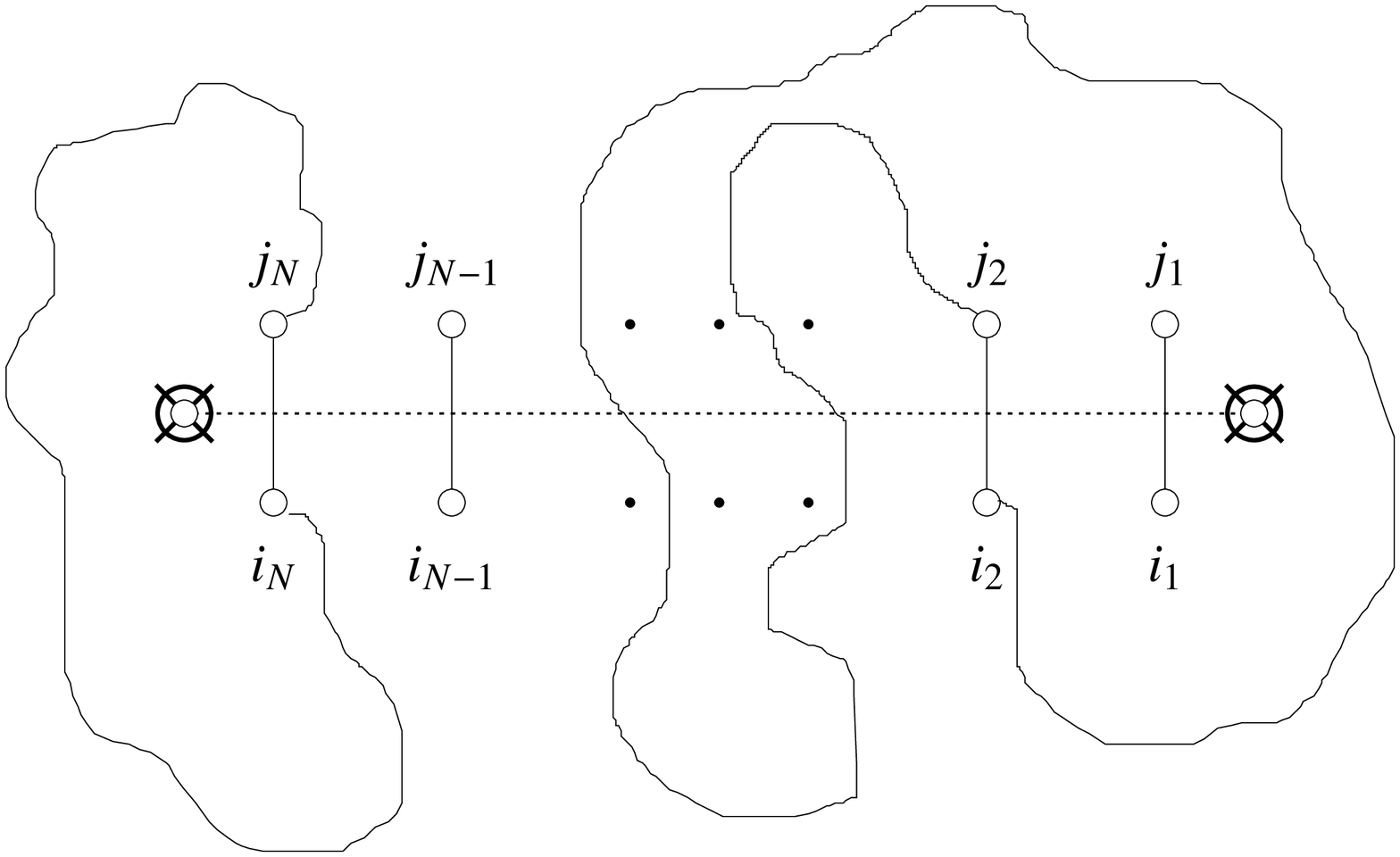}
\caption{Typical loop configurations in presence of a defect line of length $N$. The vacancy (on the left) is odd, while the sites $i_k,j_k$ are even.}
\label{slit0}
\end{figure}

Consider first the situation of a finite defect line, of length $N$.
The system is first confined in a finite grid; the infinite volume limit is then taken, while keeping $N$ finite but large.
This situation is geometrically depicted in Fig.\ref{slit0}. \vspace{-5mm}

The perturbed Laplacian corresponding to the finite defect line is $\tilde\Delta^{\rm op} = \Delta^{\rm op} + B^{(0)}$
with the non-zero part of $B^{(0)}$ given by
\renewcommand{\arraystretch}{0.7}
\arraycolsep 5pt
\begin{equation}
B^{(0)}=
\left(
\begin{array}{cccc|cccc}
 0  &  0  & \ldots &  0  &  2  &  0  & \ldots &  0 \\
 0  &  0  & \ldots &  0  &  0  &  2  & \ldots &  0 \\
\vdots&\vdots&\ddots&\vdots&\vdots&\vdots&\ddots&\vdots \\
 0  &  0  & \ldots &  0  &  0  &  0  & \ldots &  2 \\
\hline
 2  &  0  & \ldots &  0  &  0  &  0  & \ldots &  0 \\
 0  &  2  & \ldots &  0  &  0  &  0  & \ldots &  0 \\
\vdots&\vdots&\ddots&\vdots&\vdots&\vdots&\ddots&\vdots \\
 0  &  0  & \ldots &  2  &  0  &  0  & \ldots &  0 \\
\end{array}\right).
\end{equation}
The defect matrix $B^{(0)}$ is a $2N \times 2N$ matrix, with indices ranging over $i_1,\ldots,i_N,j_1,\ldots,j_N$.

We want to compute the ratio
\begin{equation}
\mathcal{D}^{(0)}_N \equiv {\det \tilde\Delta^{\rm op} \over \det \Delta^{\rm op}} = \det(I + G^{\rm op}B^{(0)}),
\label{d0}
\end{equation}
which is the fraction of arrow configurations on the even sublattice with at most $N$ loops, each containing an odd number of frustrated bonds,
compared to the number of spanning trees. In terms of dimers, we can think of two odd vacancies placed at the two ends of the defect line,
with the loops enclosing one of the two vacancies.

If the even sublattice is a finite grid of size $L$, the determinants in the numerator and denominator are $L^2 \times L^2$,
but because the matrix $B^{(0)}$ has rank $2N$, the rightmost determinant in (\ref{d0}) is only $2N \times 2N$.
When $L$ goes to infinity, the  order of this determinant remains equal to $2N$,
and the inverse Laplacian $G^{\rm op}$ may be replaced by the inverse Laplacian $G$ on the infinite lattice.
Then the simple block form of $B^{(0)}$ makes it possible to factorize the corresponding determinant into two $N$-by-$N$
determinants of Toeplitz matrices:
\begin{equation}
\mathcal{D}^{(0)}_N =
\det
\left(\begin{array}{c|c}
S & S' \\
\hline
S' & S
\end{array} \right)
= \left(\begin{array}{c|c}
S-S' & S' \\
\hline
0 & S + S'
\end{array} \right)
= \det (S - S')_N \det (S + S')_N,
\label{d0N}
\end{equation}
with the matrices $S,S'$ given by
\begin{eqnarray}
S_{k,\ell} \egal \delta_{k,\ell} + 2 G_{i_k,j_\ell} = \delta_{k,\ell} + 2 G_{j_k,i_\ell},
\label{S}\\
S'_{k,\ell} \egal 2 G_{i_k,i_\ell} = 2 G_{j_k,j_\ell},
\label{SS}
\end{eqnarray}
for $k,\ell=1,2,\ldots,N$.
The translational invariance of the Green function,
$G_{\vec r,\vec r\,'} = G_{\vec r -\vec r\,',\vec 0} \equiv G_{p,q}$ with $(p,q)=\vec r - \vec r\,'$,
shows that $S$ and $S'$ are Toeplitz matrices, and the reflection symmetries imply
\begin{equation}
S_{k,\ell} = \delta_{k,\ell} + 2G_{k-\ell,1}\,, \quad {\rm and} \quad
S'_{k,\ell} = 2G_{k-\ell,0}\,.
\label{skl}
\end{equation}
From the explicit formula,
\begin{equation}
G_{p,q} = \int_{-\pi}^{\pi} \;{{\rm d}\varphi \over 4\pi} \, {[2-\cos \varphi-\sqrt{(2 - \cos \varphi)^2-1}]^{|q|}  \over \sqrt{(2 - \cos \varphi)^2-1}} \, {\rm e}^{\i\varphi p},
\end{equation}
we find the generating functions
\begin{eqnarray}
f_+(\varphi) \iden \sum_{n=-\infty}^{+\infty} (S_n + S_n') {\rm e}^{\i n \varphi} = \sqrt{6 - 2 \cos \varphi \over 2 - 2 \cos \varphi} \label{f+}\\
f_-(\varphi) \iden \sum_{n=-\infty}^{+\infty} (S_n - S_n') {\rm e}^{\i n \varphi} = \sqrt{2 - 2 \cos \varphi \over 6 - 2 \cos \varphi} = {1 \over f_+(\varphi)}.
\label{f-}
\end{eqnarray}

The determinants of $S \pm S'$ can be computed by using Widom's theorem, a generalization of Szeg\"o's theorem \cite{widom}.
Let $f(\varphi) = \sum\limits_{n} \, a_n \, {\rm e}^{\i n \varphi}$ be a function on the unit circle of the form
\begin{equation}
f(\varphi) = (2 - 2 \cos \varphi )^{\alpha} \, g(\varphi), \qquad \alpha > -{1 \over 2},
\end{equation}
where $g(\varphi)$ is a smooth univalent function, nowhere vanishing nor divergent. Then the asymptotic value of the Toeplitz determinant formed with the Fourier coefficients $a_n$ of $f$, is given by
\begin{equation}
D_{N-1}(f) \equiv \det (a_{\ell-k})_{1 \leq k,\ell \leq N} \simeq E[g;\alpha] \, N^{\alpha^2} \, {\rm e}^{N(\log g)_0}, \qquad N \gg 1,
\end{equation}
where $(\log g)_0$ is the zeroth Fourier coefficient of $\log g$, and $E[g;\alpha]$ is a constant whose explicit value can be found in \cite{widom}.

Applying this theorem to $f_+$ and $f_-$ yields
\begin{equation}
(\log g_+)_0 = \frac{1}{2\pi} \int_{0}^{2\pi} \log g_+(\varphi) \, d\varphi =\log(1+\sqrt{2}), \qquad (\log g_-)_0 = -\log(1+\sqrt{2}),
\end{equation}
and consequently,
\begin{equation}
\mathcal{J}^{\pm}_N \equiv \det (S \pm S')_N \simeq E_\pm \, N^{\frac{1}{4}} \,\left(1+\sqrt{2}\right)^{\pm N},
\label{Jm}
\end{equation}
where $E_\pm = E[g_\pm;\mp {1 \over 2}]$.
Finally, we obtain the ratio (\ref{d0}), for large $N$:
\begin{equation}
\mathcal{D}^{(0)}_N \simeq E_+ \, E_- \, N^{1 \over 2}.
\label{webs}
\end{equation}

We should notice that $E_+$ is actually divergent, with a divergence proportional to $G_{0,0}$. This is clearly due to the fact that $f_+$ has a non-integrable singularity ($\alpha_+ = -{1 \over 2}$), which manifests itself in $\mathcal{J}^+_N$ by the constant $E[g_+;\alpha_+]$ being singular when $\alpha_+ \to -{1 \over 2}$ \cite{widom}. Physically, this is not surprising either, since there are much more (infinitely more in thermodynamic limit) spanning webs than spanning trees. However the divergence does not depend on $N$, which means that we get the same ``infinity'' for $N=1$ as for any other value of $N$. This suggests the better definition of  $\mathcal{D}^{(0)}_N$ as the relative fraction of spanning webs with at most $N$ loops with respect, not to the spanning trees, but to the spanning webs with at most one loop. The corresponding ratio, after a proper regularization (see \cite{pr} for instance), would then be well-defined. In any case, we will ignore the divergence, since only ratios of $\mathcal{D}^{(0)}_N$ are relevant for what follows.

The quantity $\mathcal{D}^{(0)}_N$ as such does not quite compare with the delocalization probability studied by Bouttier {\it et al} in \cite{Bouttier}.
We may however, cut the plane by inserting a vertical boundary passing through the sites $i_1$ and $j_1$, and consider the left half-plane only. In this way, the arrow configurations forming spanning webs with loops around the single vacancy are generated, and we recover a situation similar the one studied in \cite{Bouttier}, except that we work on a half-plane rather than on a large square. Which boundary conditions, open or closed, are imposed on the vertical boundary make no difference. The above formulas (\ref{d0N}), (\ref{S}) and (\ref{SS}) remain valid provided one uses the Green function appropriate to the chosen boundary conditions. The resulting $S,S'$ matrices in (\ref{skl}) are no longer Toeplitz but pick a Hankel piece, depending on $k+\ell$. We have nevertheless computed the corresponding determinants numerically. Our numerical results for this case show that the power $1/2$ in Eq. (\ref{webs}) is changed to $1/4$, which agrees with the exponent found in \cite{Bouttier}.

It is instructive also to compare the asymptotics (\ref{webs}) with that obtained by Fisher and Stephenson \cite{fist} for the monomer-monomer correlations. Considering a pair of monomers located at two sites separated by an odd number of bonds of the original lattice, they obtained the asymptotic correlation proportional to $N^{-1/2}$, with $N$ the distance between the monomers. In our case, the vacancies belong to the same sublattice and are therefore separated by an even number of bonds. The difference between the two asymptotics demonstrates the crucial role of the loop statistics which is quite different in these cases.


\section{Moving one step}\label{sec4}

\begin{figure}[t]
\includegraphics[height=30mm]{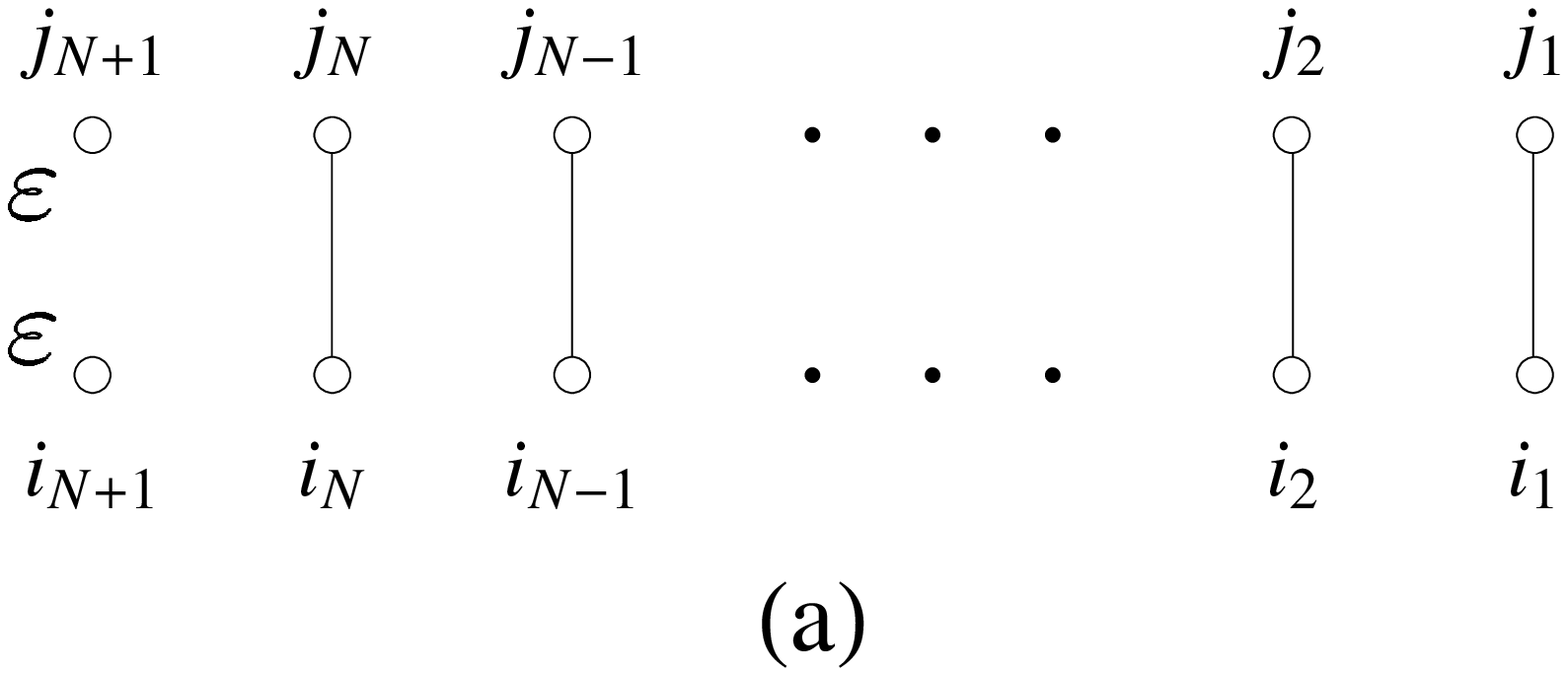}
\quad\quad\quad\quad\quad\quad
\includegraphics[height=30mm]{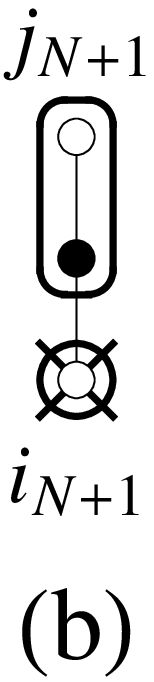}
\caption{Figure (a) describes the pertubation needed to force the presence of a dimer pointing to the vacancy, as shown in (b).}
\label{slit1}
\end{figure}

As a preparation, we start by computing the probability $\mathcal{P}_{mob}$ that the vacancy can move at least one step in a specific direction, in the infinite volume limit. Due to the rotational symmetry, $\mathcal{P}_{mob}$ can be interpreted as an efficient mobility of the vacancy in any direction. For convenience, we use in this section the spanning webs defined on the odd sublattice, so that the sites $i_k,j_k$ defining the defect line are on the same sublattice as the odd vacancy, located at the site $i_{N+1}$, see Fig.\ref{slit1}.

With respect to the discussion in Section \ref{sec2}, the difference implied by the use of the odd sublattice is that the Laplacian
of reference is now $\Delta_\otimes$, namely the Laplacian with closed boundary condition and a root at $i_{N+1}$.
For actual calculations, one can take $(\Delta_\otimes)_{ij} = (\Delta^{\rm cl})_{ij} + \varepsilon \,\delta_{i,i_{N+1}}\,\delta_{j,i_{N+1}}$
in which case the number of spanning trees growing from the root is equal to $\lim\limits_{\varepsilon \to \infty} {1 \over \varepsilon} \Delta_\otimes$.
As in the previous section, loops are included by using the perturbed Laplacian $\tilde\Delta_\otimes = \Delta_\otimes + B^{(0)}$.
The asymmetry of the descriptions afforded by the even and odd vacancies, mentioned in section II,
implies that the number of dimer configurations with one vacancy is equal
$Z_{\rm dimer} = \lim\limits_{\varepsilon \to \infty} {1 \over \varepsilon} \det \tilde\Delta_\otimes$ for an odd vacancy or
$Z_{\rm dimer} = \frac{1}{2}\det \tilde\Delta^{\rm cl}$ for an even vacancy, with $\tilde\Delta^{\rm cl} = \Delta^{\rm cl} + B^{(0)}$. On a finite grid, these two numbers are slightly different, but become equal in the thermodynamic limit.

Suppose that the vacancy can take one step upwards, so that the dimer covering the odd site $j_{N+1}$ is oriented toward $i_{N+1}$,
like on Fig.\ref{slit1}b. In terms of arrows, there is an arrow pointing from $j_{N+1}$ to $i_{N+1}$.
The number of arrow configurations with this arrow from $j_{N+1}$ to $i_{N+1}$ is equal to the number of arrow configurations
for which $j_{N+1}$ is also a root (in addition to $i_{N+1}$). Indeed the two roots generate separate branches attached to them,
but these branches can be joined to a single tree, if we place a fixed arrow (dimer) from $j_{N+1}$ to $i_{N+1}$.
In this way, we generate all spanning webs on the odd sublattice with an odd vacancy at $i_{N+1}$ and an arrow on the bond $[j_{N+1},i_{N+1}]$.
Therefore, by enumerating these spanning webs and dividing the result by the total number of spanning webs without the arrow constraint,
we obtain the probability $\mathcal{P}_{\rm mob}$ that the vacancy can make a step upwards.

Turning the site $j_{N+1}$ into a root requires to further perturb $\tilde\Delta_\otimes$ by the diagonal term $\varepsilon \,\delta_{i,j_{N+1}}\,\delta_{j,j_{N+1}}$. The resulting Laplacian can be written as $\Delta^{\rm cl} + B^{(1)}$, where the non-zero entries of $B^{(1)}$ are
\bea
B_{i_k,j_k}^{(1)} = B_{j_k,i_k}^{(1)} \egal 2\,,\qquad k=1,2,\ldots,N,\\
B_{i_{N+1},i_{N+1}}^{(1)} \egal B_{j_{N+1},j_{N+1}}^{(1)} = \varepsilon.
\eea

On a finite grid, the probability that the vacancy can make a step upwards is equal to the ratio
\begin{equation}
{\lim\limits_{\varepsilon \to \infty} \frac{1}{\varepsilon^{2}}
\det(\Delta^{\rm cl} + B^{(1)}) \over \lim\limits_{\varepsilon \to \infty} \frac{1}{\varepsilon}\det \tilde\Delta_\otimes}.
\label{one-step1}
\end{equation}
In the thermodynamic limit, the denominator can be replaced by $\frac{1}{2} \det \tilde\Delta^{\rm cl}$, and $\Delta^{\rm cl}$ by the Laplacian on the plane $\Delta$, so
that the previous ratio goes to
\begin{equation}
2{\lim\limits_{\varepsilon \to \infty} \frac{1}{\varepsilon^{2}}
\det(\Delta^{\rm cl} + B^{(1)}) \over \det(\Delta^{\rm cl} + B^{(0)})} =
2{\lim\limits_{\varepsilon \to \infty} \frac{1}{\varepsilon^{2}}
\det(\Delta + B^{(1)}) \over \det(\Delta + B^{(0)})} =
2{\lim\limits_{\varepsilon \to \infty} \frac{1}{\varepsilon^{2}} \det(I + G B^{(1)}) \over \det(I + G B^{(0)})} \equiv
2\frac{\mathcal{D}^{(1)}_N}{\mathcal{D}^{(0)}_N},
\label{one-step2}
\end{equation}
of which the limit $N \to \infty$ readily yields $\mathcal{P}_{\rm mob}$.

The nonzero part of the defect matrix $B^{(1)}$ consists of two blocks of size $N+1$. As in the previous case, the corresponding determinant $\mathcal{D}^{(1)}_N$ can be factorized into two Toeplitz matrices of size $N+1$,
\begin{equation}
\mathcal{D}^{(1)}_N = -\frac{1}{4}
\det (S - S' - \mathbf{\delta}^{(N+1)})_{N+1}
\det (S + S' - \mathbf{\delta}^{(N+1)})_{N+1},
\end{equation}
for the same matrices $S_{k,\ell}$ and $S'_{k,\ell}$ as in the previous section, except that their indices run from $1$ to $N+1$, and where the matrices $\mathbf{\delta}^{(s)}$ are defined as $\delta^{(s)}_{k,\ell}=\delta_{k,s}\delta_{\ell,s}$ for any $s=1,2,\ldots\,$. The expansion of these determinants by the last element $(N+1,N+1)$ gives
\begin{equation}
\det (S \pm S' - \delta^{(N+1)})_{N+1} = \mathcal{J}^{\pm}_{N+1} - \mathcal{J}^{\pm}_{N}.
\end{equation}
Therefore, using (\ref{Jm}), (\ref{webs}) and (\ref{one-step2}), we obtain
\begin{equation}
\mathcal{P}_{\rm mob} = \lim_{N\to\infty} \frac{2\mathcal{D}^{(1)}_N}{\mathcal{D}^{(0)}_N} = \sqrt{2} - 1.
\end{equation}						


\section{Localization probabilities of the vacancy}\label{sec5}

Dimer configurations with a vacancy on the odd sublattice are in correspondence with  spanning webs on the even sublattice.
As recalled above, the central part of an even spanning web consists of a loop surrounding a certain number of odd sites of the original lattice,
which exactly form the domain accessible to the vacancy.
Following \cite{Bouttier}, let $p_L(s)$ be the probability that this domain has size $s$ when the grid is finite (with size proportional to $L$),
and let $p(s)$ be its infinite volume limit value. Hereafter, we compute $p(1)$ and $p(2)$,
the probabilities that the vacancy can take respectively zero or one move, so that $p(1)$ is the probability that the vacancy be fully jammed.
We start with $p(1)$.

\begin{figure}[t]
\includegraphics[height=30mm]{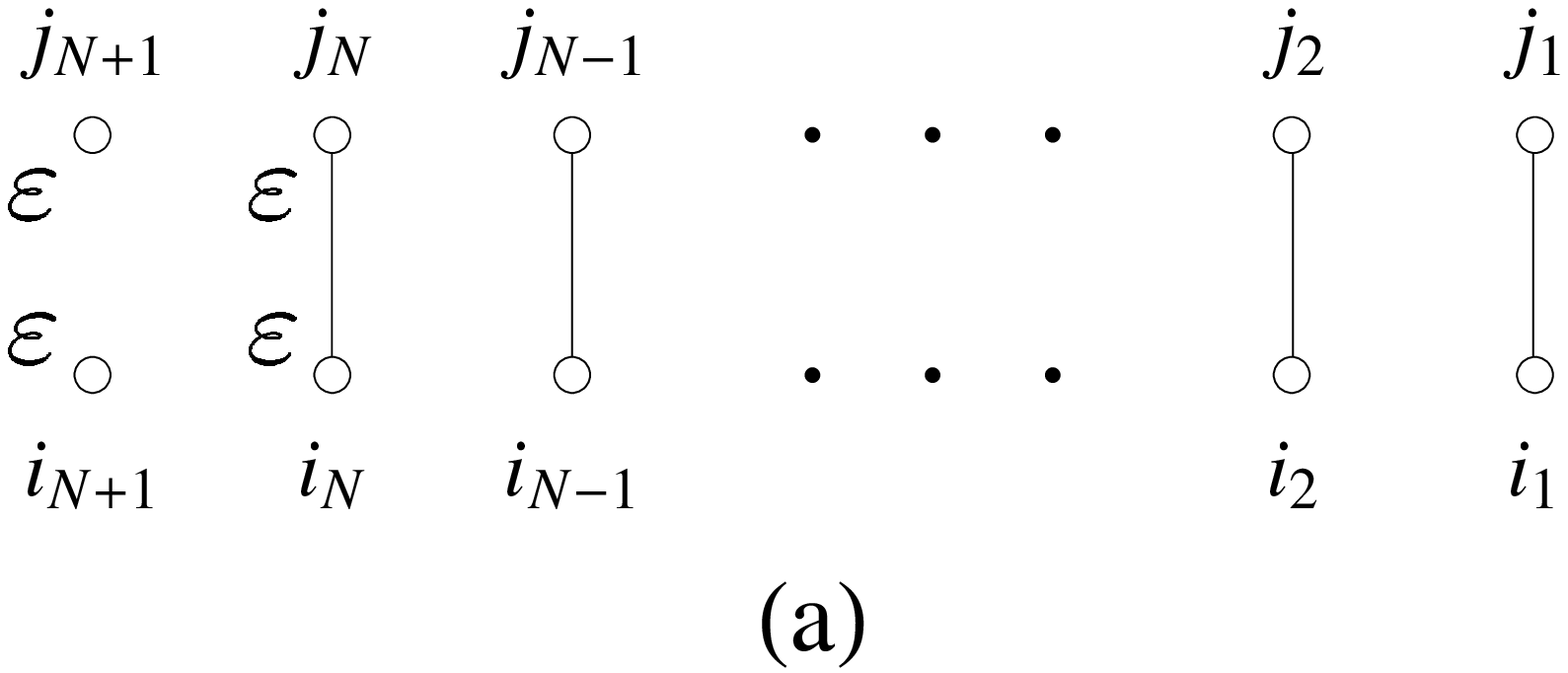}
\quad\quad\quad\quad\quad\quad
\includegraphics[height=30mm]{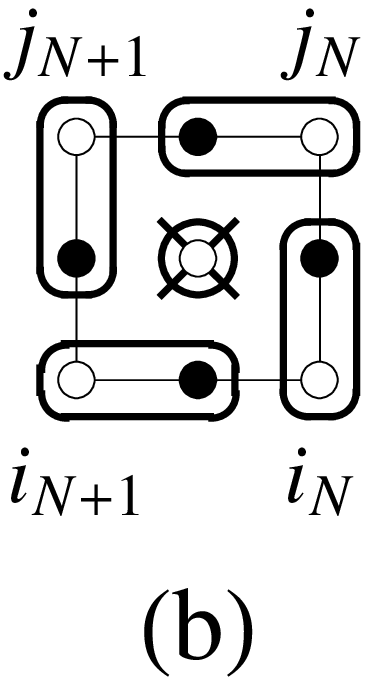}
\caption{In (a) is a pictorial description of the perturbation needed to account for the presence of a minimal loop surrounding the vacancy, as shown in (b).}
\label{slit2}
\end{figure}

If the vacancy is fully jammed, it has to be surrounded by four dimers, arranged as in Fig.\ref{slit2}b or a similar pattern in which the dimer covering $i_N$ points west.
In the two cases, the arrows drawn on the even sublattice are constrained to form a minimal loop around the vacancy,
so that the arrows originating from the four sites $i_{N},i_{N+1},j_{N},j_{N+1}$ must form a cycle.

Let us now turn the sites $i_{N},i_{N+1},j_{N},j_{N+1}$ into roots, like in the previous section. On the rest of the even sublattice, the arrows are all unconstrained, and can possible point towards $i_{N},i_{N+1},j_{N}$ or $j_{N+1}$. If in addition, we insert a defect line of length $N$ as before, the arrows may form loops passing through the bonds $(i_1,j_1), (i_2,j_2), \ldots (i_{N-1},j_{N-1})$. These changes lead to the perturbed Laplacian $\tilde\Delta_N^{\rm op} = \Delta^{\rm op} + B^{(2)}$, where the non-zero entries of the defect matrix are given by
\begin{eqnarray}
&& \hspace{-5mm} B_{i_k,j_k}^{(2)} = B_{j_k,i_k}^{(2)} = 2\,,\qquad k=1,2,\ldots,N,\\
&& \hspace{-5mm} B_{i_{N},i_{N}}^{(2)} = B_{i_{N+1},i_{N+1}}^{(2)} = B_{j_{N},j_{N}}^{(2)} = B_{j_{N+1},j_{N+1}}^{(2)} = \varepsilon,
\end{eqnarray}
for $\varepsilon$ very large.

The determinant $\lim\limits_{\varepsilon \to \infty} {1 \over
\varepsilon^4} \det \tilde\Delta_N^{\rm op}$ counts the number of even spanning webs with at most $N-1$ loops in presence of
the four roots. Every such spanning web gives rise to two dimer configurations on the whole lattice with nine vacancies, namely
the original vacancy and the eight surrounding sites. In turn these eight sites can be covered by four dimers in two ways.
Thus the number of dimer coverings in which four dimers loop around the vacancy (and with the number of even loops bounded by $N$)
is equal to $4 \times \lim\limits_{\varepsilon \to \infty} {1 \over \varepsilon^4} \det \tilde\Delta_N^{\rm op}$, and their
fraction in the set of similar dimer coverings but with no restriction around the vacancy is
\begin{equation}
4 \times {\lim\limits_{\varepsilon \to \infty} \frac{1}{\varepsilon^{4}}
\det(\Delta^{\rm op} + B^{(2)})\over \det(\Delta^{\rm op} + B^{(0)})} =
4 \times {\lim\limits_{\varepsilon \to \infty} \frac{1}{\varepsilon^{4}} \det(I + G^{\rm op} B^{(2)})\over \det(I + G^{\rm op} B^{(0)})}.
\end{equation}
In the thermodynamic limit, $G^{\rm op}$ can be replaced by the Green matrix on the plane, and the previous ratio becomes
\begin{equation}
4\;  {\mathcal{D}^{(2)}_N \over \mathcal{D}^{(0)}_N} \equiv
4 \times {\lim\limits_{\varepsilon \to \infty} \frac{1}{\varepsilon^{4}} \det(I + G B^{(2)})\over \det(I + G B^{(0)})}.
\end{equation}
Finally the limit $N \to \infty$ gives the jamming probability $p(1)$,
\begin{equation}
p(1) = 4 \, \lim_{N \to \infty} \: {\mathcal{D}^{(2)}_N \over \mathcal{D}^{(0)}_N}.
\end{equation}
where $\mathcal{D}^{(0)}_N$ is known from (\ref{webs}).

The rank $2N+2$ determinant defining $\mathcal{D}^{(2)}_N$ can be factorized as before,
\begin{equation}
\mathcal{D}^{(2)}_N = \frac{1}{16} \det (S - S' - \delta^{(N)} - \delta^{(N+1)})_{N+1} \times \det (S + S' - \delta^{(N)} - \delta^{(N+1)})_{N+1},
\end{equation}
and yields the following form for the jamming probability $p(1)$,
\begin{equation}
p(1) = {1 \over 4} \lim_{N \to \infty}
{\det (S - S' - \delta^{(N)} - \delta^{(N+1)})_{N+1} \over \det (S-S')_N}
\times {\det (S + S' - \delta^{(N)} - \delta^{(N+1)})_{N+1} \over \det(S+S')_N}.
\end{equation}

Expanding the determinants and using the values of $\mathcal{J}^{\pm}_N = \det(S \pm S')_N$ given in (\ref{Jm}) recasts the ratios as
\begin{eqnarray}
{\det (S \pm S' - \delta^{(N)} - \delta^{(N+1)})_{N+1} \over \det (S \pm S')_N} \egal
{\mathcal{J}^{\pm}_{N+1} \over \mathcal{J}^{\pm}_N}
- {{\rm Mi}_{(N;N)}(S \pm S')_{N+1} \over \det (S \pm S')_N}
- {\mathcal{J}^{\pm}_{N} \over \mathcal{J}^{\pm}_N}
+ {\mathcal{J}^{\pm}_{N-1} \over \mathcal{J}^{\pm}_N} \nonumber\\
\egal 2\sqrt{2} - 1 - {{\rm Mi}_{(N;N)}(S \pm S')_{N+1} \over \det (S \pm S')_N} + o(1),
\label{2ratios}
\end{eqnarray}
to leading order in $N$.
The notation ${\rm Mi}_{(N;N)}(S \pm S')_{N+1}$ stands for the principal minor
$\det(S_{ij} \pm S'_{ij})_{1 \leq i,j \leq N+1,i,j \neq N}$.

Minors of Toeplitz matrices have been very recently studied by Bump and Diaconis, who have obtained general formulae \cite{budi} (see also \cite{trawi}). The expressions useful for the case at hand, and for the evaluation of $p(2)$ below, are briefly recalled in the Appendix. Using them, we obtain the asymptotic value
\begin{equation}
\lim_{N \to \infty} {{\rm Mi}_{(N;N)}(S \pm S')_{N+1} \over \det (S \pm S')_N} = 1 + (\log{f_\pm})_{-1} (\log{f_\pm})_1 = 4 - 2\sqrt{2}.\label{minor1}
\end{equation}
The two ratios (\ref{2ratios}) are then both equal to $4\sqrt{2} - 5$ for large $N$, and yield the result quoted in \cite{Bouttier},
\begin{equation}
p(1) = {1 \over 4} (4 \sqrt{2} - 5)^2 = \frac{57}{4}-10\sqrt{2}.
\end{equation}

\begin{figure}[t]
\includegraphics[height=30mm]{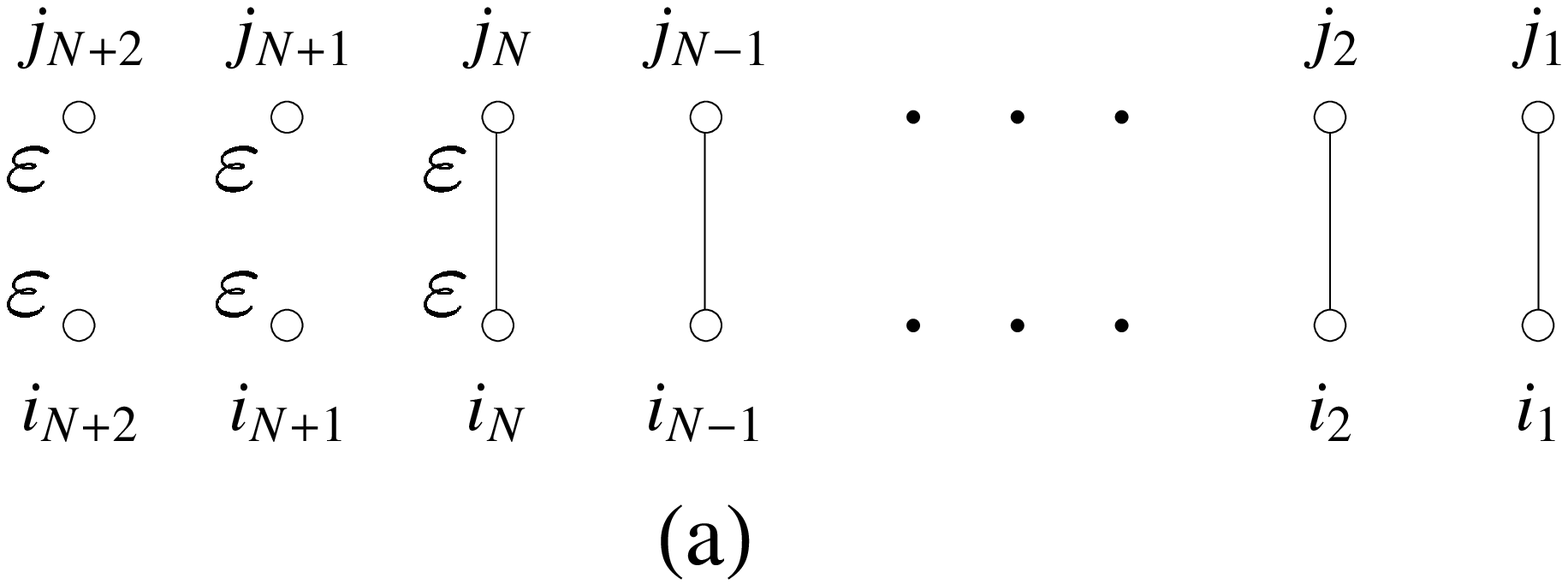}
\quad\quad\quad\quad\quad\quad
\includegraphics[height=30mm]{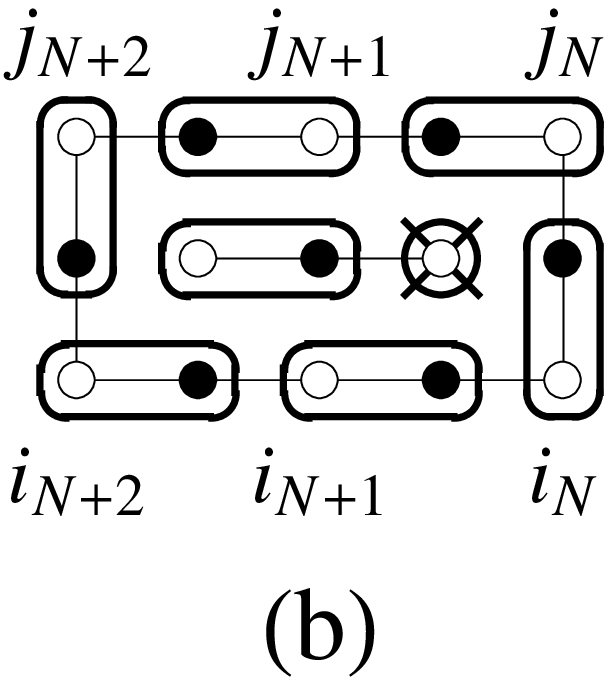}
\caption{Geometric setting for the calculation of $p(2)$, corresponding to a 6-dimer loop around the vacancy.}
\label{slit3}
\end{figure}

The similar calculation can be carried out for $p(2)$.
If the vacancy can take one and only one step, the dimers around it must be organized as
in Fig.\ref{slit3}b.
In terms of arrows drawn on the even sublattice, they all correspond to fix a rectangular length $6$ loop running around the vacancy.
The shape of the loop can be rotated by $90$ degrees, and each loop may have two orientations, making eight different situations.
They however all contribute the same number.

The arguments used for $p(1)$ apply to this case as well,
leading to a perturbation of the Laplacian involving six roots and a defect line of length $N$.
The corresponding defect matrix $B^{(3)}$ therefore reads
\begin{eqnarray}
&& \hspace{-7mm} B_{i_k,j_k}^{(3)} = B_{j_k,i_k}^{(3)} = 2\,,\qquad k=1,2,\ldots,N,\\
&& \hspace{-7mm} B_{i_{N},i_{N}}^{(3)} = B_{i_{N+1},i_{N+1}}^{(3)} = B_{i_{N+2},i_{N+2}}^{(3)} = B_{j_{N},j_{N}}^{(3)} = B_{j_{N+1},j_{N+1}}^{(3)} B_{j_{N+2},j_{N+2}}^{(3)} = \varepsilon.
\end{eqnarray}

Reasoning as before, the determinant $\lim\limits_{\varepsilon \to \infty} {1 \over \varepsilon^6} \det (\Delta^{\rm op} + B^{(3)})$
counts the number of even spanning webs with at most $N-1$ loops in presence of six roots.
Again every such spanning web gives rise to two dimer configurations on the whole lattice with fifteen vacancies, namely all the sites shown in Fig.\ref{slit3}b.
If one excepts the vacancy itself, the other fourteen sites can be covered by seven dimers in eight different ways,
among which only two contain a 6-dimer loop.
So the number of dimer coverings in which six dimers make a loop around the vacancy (and with the number of even loops bounded by $N$)
is equal to $4 \times \lim\limits_{\varepsilon \to \infty} {1 \over \varepsilon^6} \det(\Delta^{\rm op}+B^{(3)})$.
Their fraction in the set of similar dimer coverings but with no restriction around the vacancy is, in the thermodynamic limit, equal to
\begin{equation}
4\;  {\mathcal{D}^{(3)}_N \over \mathcal{D}^{(0)}_N} \equiv
4 \times {\lim\limits_{\varepsilon \to \infty} {1 \over \varepsilon^6} \det(I + G B^{(3)})\over \det(I + G B^{(0)})}.
\end{equation}
Taking the limit $N \to \infty$ and multiplying the result by 4
to account for the four rotations of the rectangle yields the jamming probability $p(2)$,
\begin{equation}
p(2) = 16 \, \lim_{N \to \infty} \: {\mathcal{D}^{(3)}_N \over \mathcal{D}^{(0)}_N}.
\end{equation}

The factorized form of $\mathcal{D}^{(3)}_N$ reads
\begin{equation}
\mathcal{D}^{(3)}_N = -\frac{1}{64} \det (S - S' - \delta^{(N)} - \delta^{(N+1)} - \delta^{(N+2)})_{N+2} \times \det (S + S' - \delta^{(N)} - \delta^{(N+1)} - \delta^{(N+2)})_{N+2}.
\end{equation}
Straightforward algebra reduces the ratios to previous results and two new types of minors,
\begin{eqnarray}
{\det (S \pm S' - \delta^{(N)} - \delta^{(N+1)} - \delta^{(N+2)})_{N+2} \over \det (S \pm S')_{N+1}} =
\left\{
\begin{matrix}
12\sqrt{2} - 16 \\
4\sqrt{2}-8
\end{matrix}\right\} - \nonumber\\
\noalign{\medskip}
- {{\rm Mi}_{(N;N)}(S \pm S')_{N+2} \over \det(S \pm S')_{N+1}}
+ (\sqrt{2} \mp 1) {{\rm Mi}_{(N,N+1;N,N+1)}(S \pm S')_{N+2} \over \det(S \pm S')_N} + o(1),
\label{minor2}
\end{eqnarray}
to leading order in $N$, and where ${\rm Mi}_{(N,N+1;N,N+1)}(S \pm S')_{N+2}$ is the determinant
$\det(S_{ij} \pm S'_{ij})_{1 \leq i,j \leq N+2, i,j \neq N,N+1}$.
The evaluation of these two minors is again a simple application of the Bump-Diaconis formulae \cite{budi}.
Surprisingly we find that the four minors are equal by pairs,
\begin{equation}
\lim_{N \to \infty}{{\rm Mi}_{(N;N)}(S \pm S')_{N+2} \over \det(S \pm S')_{N+1}} = \lim_{N \to \infty}{{\rm Mi}_{(N,N+1;N,N+1)}(S \mp S')_{N+2} \over \det(S \mp S')_N} =
\begin{cases}
{265 \over 4} - 46\sqrt{2},\\
{73 \over 4} - 12\sqrt{2}.
\end{cases}
\label{minor3}
\end{equation}

Inserting these results in the above equations, we obtain the exact value for $p(2)$,
\begin{equation}
p(2) = \frac{1}{32}(72\,817 \sqrt{2} - 102\,977),
\end{equation}
as it was quoted in the Introduction.


\section*{Acknowledgments}
This work was supported by RFBR grant No 06-01-00191a, and by the Belgian Interuniversity Attraction Poles Program P6/02. P.R. is a Research Associate of the Belgian National Fund for Scientific Research (FNRS).

\def\theequation{A.\arabic{equation}}

\appendix*
\section{Minors of Toeplitz matrices}

We collect here the various formulae used in Section \ref{sec5} to compute specific minors of Toeplitz matrices. These are but particular cases of more general expressions proved by Bump and Diaconis in \cite{budi}. In fact the asymptotic values of exactly the same determinants have been obtained by Tracy and Widom \cite{trawi} at about the same time. However their general answer takes a different form and appears to be less convenient for concrete calculations.

Generically these formulae evaluate the determinants of Toeplitz matrices with certain rows and columns removed or shifted, and which are, as a consequence, no longer Toeplitz. For $f(\varphi) = \sum\limits_{n} \, a_n \, {\rm e}^{\i n \varphi}$ a function on the unit circle, let $T_{N-1}(f) = (a_{\ell-k})_{1 \leq k,\ell \leq N}$ be the usual rank $N$ Toeplitz matrix associated to the symbol $f$, and $D_{N-1}(f) = \det T_{N-1}(f)$ its determinant.

Now let $\lambda=(\lambda_1,\lambda_2,\lambda_3,\ldots)$ be a partition of $m$, that is, a decreasing sequence of non-negative integers $\lambda_1 \geq \lambda_2 \geq \lambda_3 \geq \ldots$ such that $\sum\limits_{k} \, \lambda_k = m$, and likewise, let $\mu=(\mu_1,\mu_2,\mu_3,\ldots)$ be a partition of $p$. For fixed $\lambda,\mu$, the determinants of interest are given by
\begin{equation}
D_{N-1}^{\lambda,\mu}(f) \equiv \det(a_{\lambda_k-\mu_\ell-k+\ell})_{1 \leq k,\ell \leq N}.
\end{equation}
For general $\lambda,\mu$, the determinant $D_{N-1}^{\lambda,\mu}(f)$ is not the minor of a larger Toeplitz matrix $T^{}_M(f)$, but a minor with certain row and column indices shifted. For example if $\lambda=(2,1,0,\ldots)$ and $\mu=(1,0,\ldots)$, $D_{N-1}^{\lambda,\mu}(f)$ is the determinant of $T^{}_N(f)$ from which we cross the third row and the second column, and then shift by 1 the indices of the Fourier coefficients lying on the first row. When $\lambda_1 = \mu_1$, $D_{N-1}^{\lambda,\mu}(f)$ can be seen as a minor of the larger matrix $T_{N-1+\lambda_1}(f)$ from which $\lambda_1$ rows and $\mu_1$ columns are crossed out.

The asymptotic values of the ratios $D_{N-1}^{\lambda,\mu}(f)/D_{N-1}^{}(f)$ for large $N$ but fixed $\lambda$ and $\mu$ have been computed by Bump and Diaconis in \cite{budi}. Their expressions look complicated but are completely explicit and simple enough for small values of $m$ and $p$. For two permutations $\pi$ and $\rho$, the numbers of $k$-cycles in $\pi$ and in $\rho$ are denoted respectively $\gamma_k$ and $\delta_k$. If $\log f(\varphi) = \sum\limits_{n} \, c_n \, {\rm e}^{\i n \varphi}$, one defines
\begin{equation}
\Delta(f,\pi,\rho) = \prod_{k=1}^\infty \;
\begin{cases}
k^{\gamma_k} \, c_k^{\gamma_k-\delta_k} \, \delta_k! \, L_{\delta_k}^{(\gamma_k-\delta_k)}(-kc_kc_{-k}) & \text{if $\gamma_k \geq \delta_k$},\\
k^{\delta_k} \, c_{-k}^{\delta_k-\gamma_k} \, \gamma_k! \, L_{\gamma_k}^{(\delta_k-\gamma_k)}(-kc_kc_{-k}) & \text{if $\delta_k \geq \gamma_k$},
\end{cases}
\end{equation}
in terms of the Laguerre polynomials
\begin{equation}
L_n^{(\alpha)}(t) = \sum_{k=0}^n \; {n+\alpha \choose n-k} \, {(-t)^k \over k!}.
\end{equation}
Then one of the main results of \cite{budi} asserts that, under suitable conditions on $f$,
\begin{equation}
\lim_{N \to \infty} {D_{N-1}^{\lambda,\mu}(f) \over D_{N-1}^{}(f)} = {1 \over m! \, p!} \;\sum_{\pi \in S_m} \: \sum_{\rho \in S_p} \; \chi^\lambda(\pi) \, \chi^\mu(\rho) \, \Delta(f,\pi,\rho),
\label{main}
\end{equation}
where $\chi^\lambda(\pi)$ is the irreducible character of $S_m$ associated to the partition $\lambda$, evaluated at the group element $\pi$, and similarly for $\chi^\mu(\rho)$.

The previous result was proved in \cite{budi} under the conditions that the Fourier coefficients of $\log f(\varphi)$ satisfy $\sum\limits_{n} |c_n| < +\infty$ and $\sum\limits_{n} |n| \cdot |c_n|^2 < +\infty$. For the applications we want to make of this result, the function $f$ should be either $f_+$ or $f_-$, given in (\ref{f+}) and (\ref{f-}), and none of them satisfies these conditions, because the factor $(2 - 2\cos\varphi)^{\mp 1/2}$ has a root or a singularity on the unit circle. However the result (\ref{main}), written in the form given above, should hold even if these conditions are not satisfied since the right-hand side only depends on a finite number of Fourier coefficients of $\log f$.

Alternatively, and since the ratio is well-defined independently of the above two conditions on $\log f$, one can circumvent the difficulty by regularizing the symbols $f_+$ and $f_-$, by defining
\begin{equation}
f_\pm(\varphi;t) = \left({6 - 2\cos{\varphi} \over 2t - 2\cos{\varphi}}\right)^{\pm 1/2}.
\end{equation}
For $t>1$, the two conditions are satisfied so that the formula (\ref{main}) can be used. Since the right-hand side is continuous at $t=1^+$, one can take the limit $t \to 1$ at the end of the calculation, or equivalently use the Fourier coefficients $c_k$ for the unregularized symbols $f_\pm$.

Let us now make contact with the minors needed in Section \ref{sec5}. The first ones are given in (\ref{minor1}), namely ${{\rm Mi}_{(N;N)}(S \pm S')_{N+1}/\det (S \pm S')_N}$. In the present notation, they are equal to the minor ${\rm Mi}_{(N;N)}(T_N(f))$ divided by $D^{}_{N-1}(f)$ for $f = f_\pm$. The matrix $T_N(f)$ being Toeplitz of order $N+1$, its $(N,N)$-minor is equal to its $(2,2)$-minor, itself equal to $D_{N-1}^{\lambda,\mu}(f)$ for the partitions $\lambda = \mu = (1)$ of $m=p=1$. The formula (\ref{main}) yields
\begin{equation}
\lim_{N \to \infty} {{\rm Mi}_{(N;N)}(T_N(f)) \over D^{}_{N-1}(f)} = \lim_{N \to \infty} {{\rm Mi}_{(2;2)}(T_N(f)) \over D^{}_{N-1}(f)} = \lim_{N \to \infty} {D_{N-1}^{\lambda,\mu}(f)\over D^{}_{N-1}(f)} = L_1^{(0)}(-c_1c_{-1}) = 1 + c_1c_{-1}.
\end{equation}
Taking $f=f_\pm$ and using $(\log f_+)_{\pm 1} = \sqrt{2}-1 = -(\log f_-)_{\pm 1}$ yields the result given in (\ref{minor1}).

The other minors we needed to compute are ${{\rm Mi}_{(N;N)}(S \pm S')_{N+2}/\det (S \pm S')_{N+1}}$ in (\ref{minor2}). The denominator is $D_N(f)$ and the numerator, equal to ${\rm Mi}_{(3;3)}(S \pm S')_{N+2}$ is nothing but $D_N^{\lambda,\mu}(f)$, again for $f=f_\pm$, and for the partitions $\lambda=\mu=(1,1)$ of 2. These two partitions label the alternating representation of $S_2$, so that the above formula yields
\begin{equation}
\lim_{N \to \infty} {D_N^{\lambda,\mu}(f)\over D^{}_N(f)} = {1 \over 2} L_2^{(0)}(-c_1c_{-1}) + {1 \over 2} L_1^{(0)}(-2c_2c_{-2}) - {1 \over 2} (c_1^2 c_{-2} + c_{-1}^2c_2) L_0^{(2)}(-c_1c_{-1}) L_0^{(1)}(-2c_2c_{-2}).
\end{equation}
Taking $f = f_{\pm}$ and using the coefficients $(\log f_+)_{\pm 1} = -(\log f_-)_{\pm 1}$ given above as well as $(\log f_+)_{\pm 2} = 3\sqrt{2}-4 = -(\log f_-)_{\pm 2}$ yields the first part of (\ref{minor3}).

The last minors to be evaluated in (\ref{minor2}) are ${\rm Mi}_{(N,N+1;N,N+1)}(S \pm S')_{N+2}/\det(S \pm S')_N$. It is not difficult to see that these ratios are equal to $D_{N-1}^{\lambda,\mu}(f)/D^{}_{N-1}(f)$ for $f=f\pm$ and for $\lambda=\mu=(2)$. Since these partitions label the trivial representation of $S_2$, the formula (\ref{main}) shows that this ratio is equal to the previous one in which all terms are taken positively, leading to
\bea
\lim_{N \to \infty} {D_{N-1}^{\lambda,\mu}(f)\over D^{}_{N-1}(f)} && \hspace{-5mm} = \nonumber\\
&& \hspace{-3.0cm}{1 \over 2} L_2^{(0)}(-c_1c_{-1}) + {1 \over 2} L_1^{(0)}(-2c_2c_{-2}) + {1 \over 2} (c_1^2 c_{-2} + c_{-1}^2c_2) L_0^{(2)}(-c_1c_{-1}) L_0^{(1)}(-2c_2c_{-2}).
\eea
The values of the Fourier coefficients reproduce the second part of (\ref{minor3}).


\end{document}